\documentclass{pasj00}
\draft

\usepackage{color}
\usepackage{ulem}

\begin{document}
\SetRunningHead{Y. Ezoe et al.}{
Time Variability of the Geocoronal Solar Wind Charge Exchange 
in the Direction of the Celestial Equator
}
\Received{}%{yyyy/mm/dd}
\Accepted{}%{yyyy/mm/dd}

\title{
Time Variability of the Geocoronal Solar Wind Charge Exchange 
in the Direction of the Celestial Equator
}

\author{
  Yuichiro \textsc{Ezoe}\altaffilmark{1},
  Ken \textsc{Ebisawa}\altaffilmark{2},
  Noriko Y. \textsc{Yamasaki}\altaffilmark{2}, 
  Kazuhisa \textsc{Mitsuda}\altaffilmark{2},\\
  Hiroshi \textsc{Yoshitake}\altaffilmark{2}, 
  Naoki \textsc{Terada}\altaffilmark{3}, 
  Yoshizumi \textsc{Miyoshi}\altaffilmark{4}, 
  and Ryuichi \textsc{Fujimoto}\altaffilmark{5} 
}

\altaffiltext{1}{
  Tokyo Metropolitan University, 1-1, Minami-Osawa, Hachioji,
  Tokyo, 192-0397, JAPAN
}

\altaffiltext{2}{
  The Institute of Space and Astronautical Science (ISAS),
  Japan Aerospace and eXpoloration Agency (JAXA),               
  3-1-1 Yoshinodai, Sagamihara, Kanagawa 229-8510, JAPAN
}

\altaffiltext{3}{
  Tohoku University, 
  6-3, Aoba, Aramakiaza, Sendai, Miyagi 980-8578, JAPAN
}

\altaffiltext{4}{
  Nagoya University, 
  Furo-cho, Chikusa-ku, Nagoya 464-8601, JAPAN
}

\altaffiltext{5}{
  Kanazawa University, 
  Kakuma-chou, Kanazawa, Ishikawa 920-1192, JAPAN
}

\email{ezoe@phys.metro-u.ac.jp}


\maketitle

\begin{abstract}
We report the detection of a time variable O\emissiontype{VII} line emission in a deep 100 ks Suzaku X-ray Imaging
Spectrometer spectrum of the Galactic Ridge X-ray emission. The observed line intensity is too strong
(11$\pm$2 line unit or photon cm$^{-2}$ s$^{-1}$ str$^{-1}$) to be emitted inside the heavily obscured
Galactic disk. It showed a factor of two time variation which shows a significant ($\sim4\sigma$) correlation 
with the solar wind O$^{7+}$ ion flux. The high line intensity and the good time correlation with the 
solar wind strongly suggests that it originated from geocoronal solar wind charge exchange emission. 
We discuss the X-ray line intensity considering a line of sight direction and also theoretical distribution 
models of the neutral hydrogen and solar wind around the Earth. Our results indicate that X-ray observations
of geocoronal solar wind charge exchange emission can be used to constrain these models.
\end{abstract}

\section{Introduction}
\label{sec:intro}

The solar wind charge exchange (SWCX) emission has been recognized
to reproduce X-ray emission in many solar system environments 
such as comets, geocorona, heliosphere, and planetary atmospheres 
(see \cite{Bhardwaj07} for review).
When an ion in the solar wind interacts with a neutral atom, 
it strips an electron(s) from the atom and X-ray or ultra-violet 
photon(s) are released as the electron relaxes into the ground state.
Therefore, the SWCX spectrum is characterized by strong emission 
lines from highly ionized atoms such as O$^{7+}$ or C$^{5+}$.
It can be distinguished from ordinary thermally equilibrium plasma
emission by strong emission lines from large principle quantum 
number states such as $n=$3--6, and strong forbidden lines.
The classical over-the barrier model \citep{Ryufuku80,Mann81}
can explain the large $n$ line feature.
Because the heavy solar wind ions are highly charged, 
their ionization potentials are larger than those of 
the neutral target.
Therefore, when the electron moves from the target to the ion, 
it transfers into an excited state of the ion whose potential
most closely matches that of the neutral. 
Consequently, the electron will enter into a high $n$ state
and then emit a strong emission line by the transition.
The strong forbidden line feature is explained by a 
statistical-distribution model \citep{Krasnopolsky04}. 
%%
%%In this model, we assume that relative population of spin 
%%triplet and singlet states is 3:1. Then, according to the 
%%transition, the strong forbidden line relative to the 
%%ordinary plasma is observed. 

Among many classes of SWCX objects in the solar system, the geocorona
is now considered as an important contamination. The low density 
neutral hydrogen around the Earth or the geocorona emits X-ray emission 
lines such as O\emissiontype{VII}, O\emissiontype{VIII}, and C\emissiontype{V} K$_\alpha$ by the SWCX.
When we would like to observe soft X-rays from objects such as clusters 
of galaxies, this component can be a contamination source. 
At the same time, SWCX X-ray emission can provide a diagnostic of the spatial 
distribution of the geocorona and solar wind near the Earth.

During the ROSAT all-sky survey, \citet{Snowden94} found unknown
long term enhancements (LTE) in the soft X-ray background. 
Motivated by the discovery of cometary X-rays \citep{Lisse96,Dennerl97}, 
\citet{Cox98} suggested that the SWCX due to geocorona can explain 
a part of the LTE. \citet{Robertson03} constructed a model to 
simulate a spatial distribution of the geocoronal SWCX emission 
using the Earth's exosphere and solar wind distributions.

Observational results, which supports these predictions, have been 
taken with Chandra, XMM-Newton, and Suzaku.
\citet{Wargelin04} found a signature of an oxygen emission line 
in the Chandra Moon observation. Because the Moon occults
any background sources, it must have originated from the geocoronal 
SWCX.
\citet{Snowden04} discovered a significant enhancement of the 
soft X-ray background during one of XMM-Newton's observations
of the Hubble Deep Field-North. The emission showed lines 
from O, Ne and Mg and also was time variable.
This phenomena was nicely explained by XMM-Newton's changing 
viewing geometry with respect to the geocorona, whose SWCX
is predicted to have a specific distribution around the 
Earth \citep{Robertson03}.

\citet{Carter08} systematically investigated XMM-Newton archival
data focusing on time variability of the soft X-ray background. 
Using solar wind data and considering the XMM-Newton's 
line of sight, similar to \citet{Snowden04}, they found that, 
in several cases, the SWCX enhancement occurred when XMM-Newton 
was on the sub-solar side of the magnetosheath which is 
predicted by \citet{Robertson03} to be a high SWCX flux region.
The SWCX-like emission was found for other cases when XMM-Newton 
had a line of sight which did not intersect the high flux region. 
They considered that these probably arise from a non geocoronal 
origin such as a coronal mass ejection (CME) passing through the 
heliosphere.

CMEs have been used to explain diffuse X-ray background in other
X-ray observations. For example, \citet{Carter10} recently reported 
a diffuse variable X-ray emission detected with XMM-Newton. They 
analyzed its spectrum and light curve in combination with the solar 
wind proton data, and concluded that this is associated with a CME 
which interacted with neutrals in the Earth's exosphere. 

Thanks to a good energy response of the CCDs onboard Suzaku, 
\citet{Fujimoto07} discovered a firm line of evidence for 
the geocoronal SWCX emission in the north ecliptic pole 
observation (RA$=272.800^\circ$, Dec$=66.000^\circ$, J2000).
Although the observation was originally conducted to take the 
sky X-ray background, they found a variability in X-ray CCD 
light curves, which apparently correlates with the solar wind 
proton flux.
They divided the data into two periods, i.e., the stable and 
flare durations and successfully found enhancement of emission 
lines such as the C\emissiontype{VI} 4p to 1s line and the O\emissiontype{VII} forbidden line, 
in the flare spectrum. 
To explain short-term variations of less than 1 hr, they 
introduced a new parameter, the point where the line of 
sight encounters the magnetosheath (see figure 8 in 
\cite{Fujimoto07}).
They concluded that the short-term X-ray variability 
have anti-correlation with the distance to the magnetosheath. 

In this paper, we studied a possible SWCX oxygen line emission 
discovered in the long Suzaku observation of the Galactic ridge 
(RA$=281.000^\circ$, Dec$=-4.070^\circ$, J2000), 
($l=28.46^\circ$, $b=-0.20^\circ$).
In \citet{Ebisawa08}, we found strong O\emissiontype{VII} emission 
which is difficult to attribute to the Galactic ridge
X-ray emission. 
Although we thought that it can be the SWCX emission, investigations
are not enough to conclude that. In this paper, to confirm this 
inspection, we conducted a timing analysis by combining the Suzaku 
data with the solar wind data.
While the north ecliptic pole observation was in the direction of 
the north pole, the Galactic ridge is near the celestial equator
and the sub-solar side of the magnetosheath, where the geocoronal 
SWCX is expected to be strong. 
The good statistics 100 ks Suzaku data provides us with a good 
opportunity to characterize the geocoronal SWCX emission.

\section{Observation}
\label{sec:obs}

The Suzaku observation of the Galactic Ridge direction was conducted
from 2005 October 28, 02:24 to October 30, 21:30 for 100 ks. 
October 28 and 30 corresponds to the Day of Year (DOY) in 2005 of 301 
and 303, respectively.
Suzaku \citep{Mitsuda07} is the fifth Japanese X-ray astronomical 
satellite which carries four X-ray CCDs (X-ray Imaging Spectrometer, 
XIS: \cite{Koyama07}). 
Due to the low-earth orbit and the large effective area, the 
back-illuminated type CCD (XIS1) has one of the lowest relative particle 
backgrounds among all X-ray CCDs in currently available X-ray observatories. 
Furthermore, the XIS1 has good energy resolution and superior 
low-energy response with negligible low-energy tails. 
These two characteristics make Suzaku ideal for studying diffuse soft X-ray emission
like the geocoronal SWCX.

\begin{figure}[p]
  \begin{center}
    \FigureFile(\textwidth,){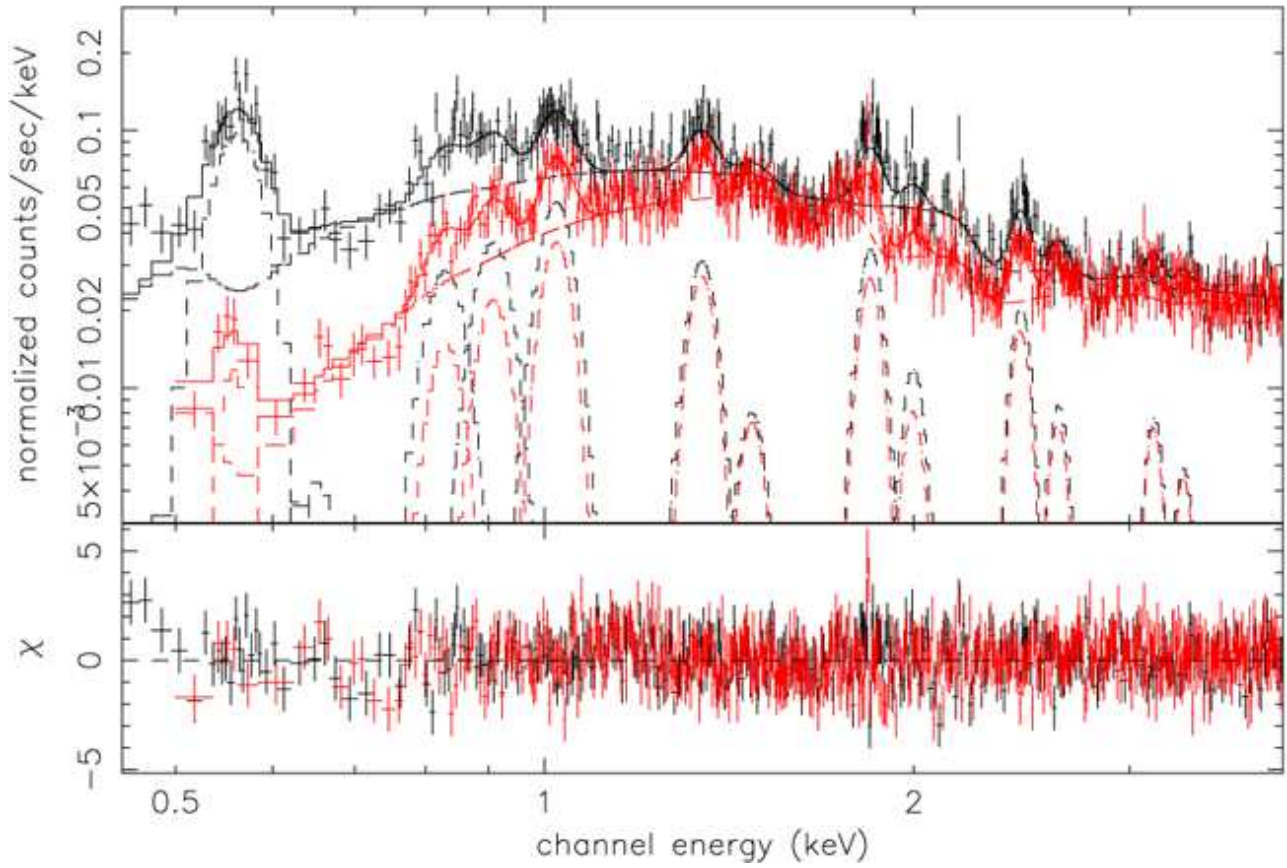}
  \end{center}
  \caption{Background-subtracted XIS1 (black) and 
    XIS$0+2+3$ (red) spectra of the Galactic ridge region.
    The solid line is the best-fit power-law plus thirteen 
    Gaussian model. The dashed lines are the model components.
    Parameters are described in table 2 of \citet{Ebisawa08}. 
    The bottom panel indicates residuals of the data from the model.
  }\label{fig:ebispec}
\end{figure}

Figure \ref{fig:ebispec} shows the XIS spectra in the low energy band 
taken from \citet{Ebisawa08}.
Among many emission lines such as Fe\emissiontype{XVII} (0.829$\pm$0.004 keV), 
Ne\emissiontype{X} (1.021$\pm$0.002 keV), Mg\emissiontype{XI} (1.342$\pm$0.003 keV), and 
Si\emissiontype{XIII} (1.843$\pm$0.003 keV), we can notice a strong
O\emissiontype{VII} line at 0.560$\pm0.03$ keV. These energies are the 
best fit values in table 2 of \citet{Ebisawa08}.
The line center energy of O\emissiontype{VII} is more likely the forbidden line (561 eV)
than the resonance line (574 eV).
Its intensity is too strong to arise from the Galactic ridge X-ray 
emission which must suffer from the interstellar hydrogen column 
density of $N_{\rm H}\sim1\times10^{21}$ 
cm$^2$.
The flux is (320$\pm$50)$\times10^{-5}$ photons s$^{-1}$ cm$^{-2}$ 
deg$^{-2}$, which corresponds to 11$\pm$2 line units (LU, photons 
s$^{-1}$ cm$^{-2}$ str$^{-1}$).
This is even larger than that of the O\emissiontype{VII} line due to SWCX found 
in the north ecliptic pole observation (5.1$^{+1.1}_{-1.0}$ LU, 
\cite{Fujimoto07}), 

This strong O\emissiontype{VII} emission line motivated us to investigate the light
curve in detail. Below we use only the back-illuminated CCD (XIS1) data 
since it has the higher statistics for the oxygen line compared to the 
front-illuminated CCDs as shown in figure \ref{fig:ebispec}.
The data reduction was performed in the same way as \citet{Ebisawa08}
except that the HEAsoft analysis package was ver 6.1.1.

\section{Timing Analysis}
\label{sec:time}

\begin{figure}[p]
  \begin{center}
    \FigureFile(\textwidth,){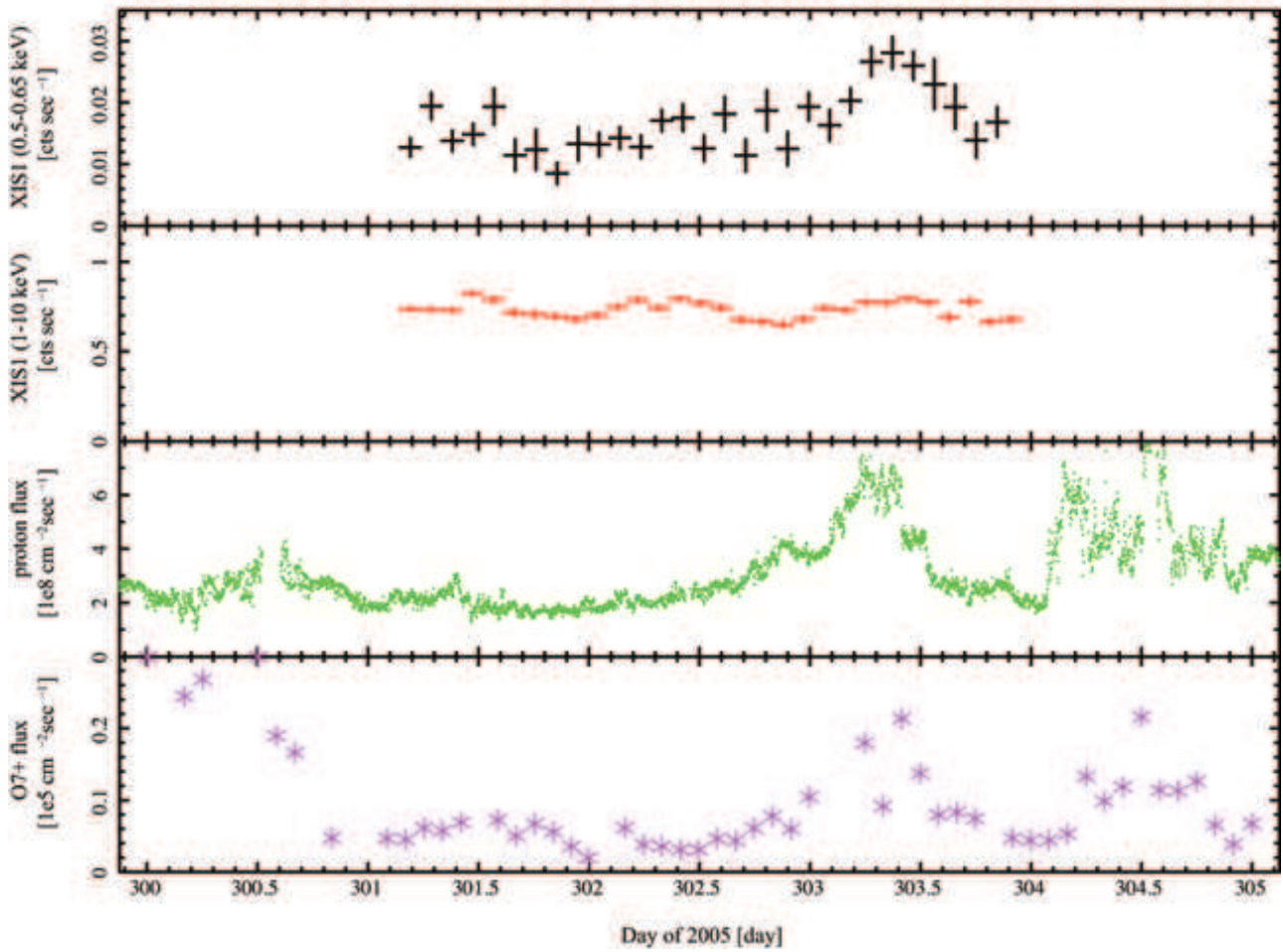}
  \end{center}
  \caption{XIS1 light curves in 0.5--0.65 keV and 1--10 keV, 
  solar wind proton flux, and O$^{7+}$ flux
  as a function of day of year (DOY) in 2005. The DOY of 301 corresponds to 2005 October 28.
  X-ray photons from calibration sources are excluded for the XIS1 1--10 keV light curve.
  The vertical error bars are at 1$\sigma$ significance.
  The solar wind proton flux is calculated from WIND SWE data (0.001 day$=$86.4 sec average), 
  while the O$^{7+}$ flux is from level 2 ACE SWICS data (2 hr average, corresponding
  to the minimum time bin). Only good data with quality flag 0 were used for the ACE data.
  }\label{fig:ovii-lc}
\end{figure}

To examine the SWCX interpretation, we created the XIS1 light curve 
in 0.5--0.65 keV where the O\emissiontype{VII} line dominates and compared it with
solar wind data. Similar to \citet{Ebisawa08}, we excluded two
point sources within the field of view (Suzaku J1844$-$0404 and 
G28.6$-$1), in order to avoid contamination from these sources.
The total area after removing two regions is 275 arcmin$^2$.
Figure \ref{fig:ovii-lc} shows the XIS1 0.5--0.65 keV light curve in 
8192 s bins, compared to the WIND proton and ACE O$^{7+}$ ion fluxes
\footnote{The WIND and ACE data were taken from ftp://space.mit.edu/pub/plasma/wind/kp\_files/ and http://www.srl.caltech.edu/ACE/ASC/level2/index.html, respectively.}. 
The position of WIND during the observation was $+$200 $R_{\rm E}$ (Earth radius)
and $-$50 $R_{\rm E}$ in the GSE x-y coordinates (i.e., at the pre-bowshock 
position\footnote{http://cdaweb.gsfc.nasa.gov/cgi-bin/gif\_walk}), 
while ACE orbited around the Lagrangian point L$_1$ between
the Sun and the Earth.
Because the ACE level 2 (publication-quality) proton data was unavailable 
during a part of the observation, we used the WIND data for the 
proton flux. The average XIS1 count rate was $1.5\times10^{-2}$ 
cts s$^{-1}$.

We can see an increase of the XIS1 light curve from 303 to 304 day. 
We checked the XIS1 light curves in different energy bands (0.65--1 
keV and 1--10 keV) but both of them did not show such a variability
(see fig. \ref{fig:ovii-lc}).
Hence, this feature seems to be intrinsic for the O\emissiontype{VII} line emission.
A similar enhancement can be seen in the solar wind proton and O$^{7+}$ fluxes.
Since the O$^{7+}$ ion is the source for the SWCX O\emissiontype{VII} line which is from
O$^{6+}$, this provides a strong line of evidence that the O\emissiontype{VII} emission 
is truly arising from the SWCX. 
The heliospheric origin can be rejected since the heliosphere is 
far larger than the Earth's geocorona in size and the short time 
variability of the solar wind will be smeared. Therefore, we conclude
that the O\emissiontype{VII} line is likely from the geocoronal SWCX.

Because the ACE satellite orbits at the Lagrangian point L$_1$ and Suzaku 
is in the low Earth orbit, we can expect $\sim1$ hr time delay between 
the XIS1 and ACE data. In order to examine the time delay, we conducted a 
cross-correlation analysis. 
This procedure needs that both light curves are taken in the equally-spaced 
time intervals. 
Since the time bins of ACE O$^{7+}$ flux and XIS1 light curve are different,
we interpolated the ACE data to match the XIS1. Considering the 
ACE O$^{7+}$ flux is a 2 hr average, both the XIS and ACE data were binned 
into 8192 s. 
We then utilized {\tt crosscor} in the HEAsoft analysis package, to obtain 
the cross correlation. In this software, we can choose several parameters 
for the calculation of the cross correlation function. We chose the default
mathematical algorithm ({\tt fast=1}, fast Fourier transform) and 
normalization method ({\tt normalization=1}, no renormalization).

\begin{figure}[p]
  \begin{center}
    \FigureFile(\textwidth,){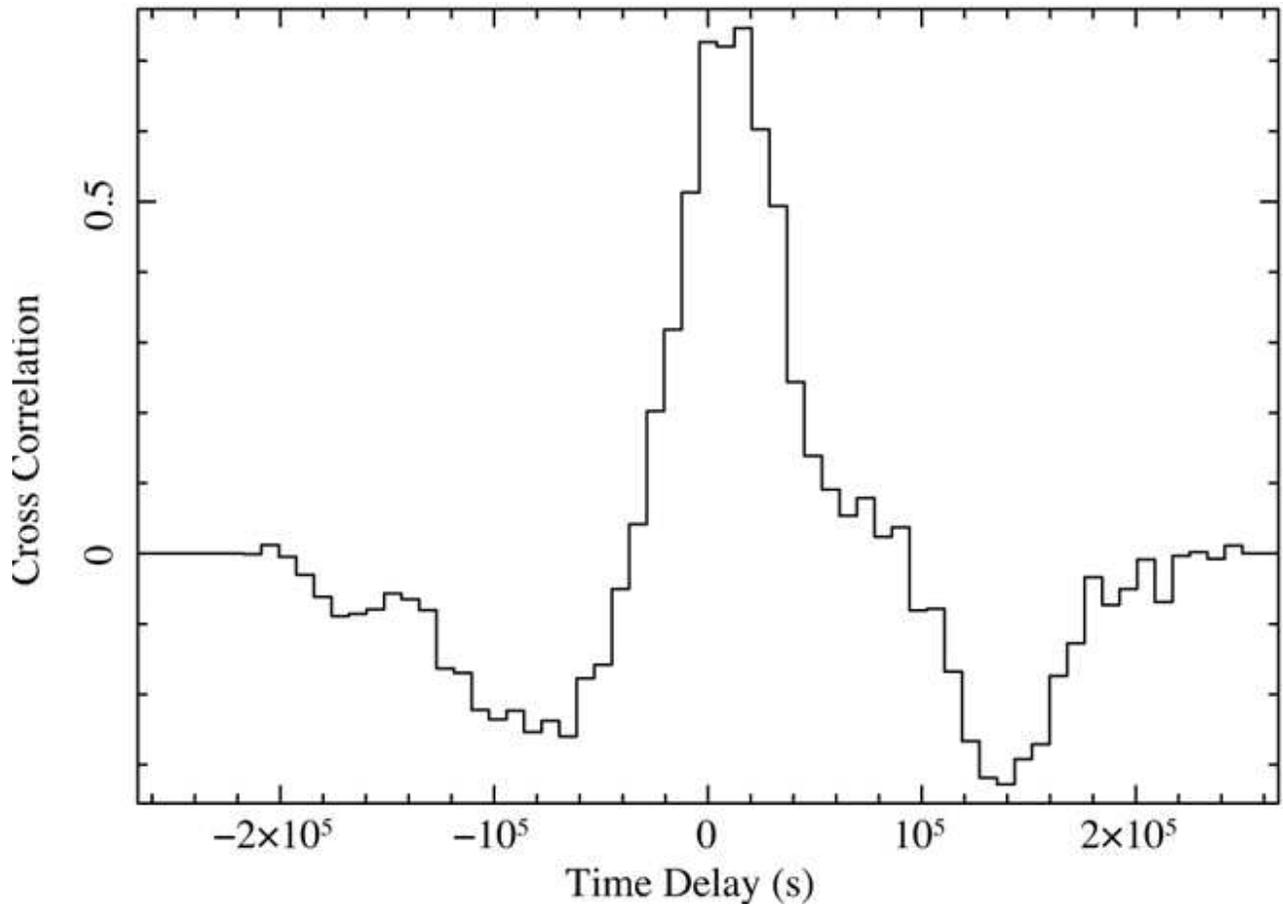}
  \end{center}
  \caption{Cross correlation between the XIS1 O\emissiontype{VII} line emission 
  and ACE O$^{7+}$ flux curves. A positive time delay means that 
  the ACE data leads the XIS1. 
  }\label{fig:crosscorr}
\end{figure}

The calculated cross correlation is shown in figure \ref{fig:crosscorr}. 
At the time delay around 0$\sim$16384 s, the correlation coefficient was 
$\sim$0.72 with the null hypothesis probability of $1\times10^{-4}$ 
corresponding to $\sim4\sigma$ significance. 
Hence, the correlations around the time delay of 0$\sim$16384 s are 
highly significant. 
We noticed that the peak is slightly shifted to the positive delay side, 
which means that the ACE data has a time delay against the XIS. 
This coincides with the fact that the ACE satellite orbits at the 
L$_1$ point and hence detects the solar wind before Suzaku. 
The expected time delay roughly depends on the distance between the 
two satellite ($\sim1.5\times10^{6}$ km) and the average solar wind 
proton speed at the observation time ($\sim400$ km s$^{-1}$). Then, 
the time delay is estimated as $\sim$3800 s. This is consistent with 
the observed time delay of 0$\sim$16384 s. 
The sparse ACE flux data hindered us to investigate a more accurate 
determination. Below we simply assume the time delay of 8192 s, 
corresponding to the peak of the cross correlation. 

\begin{figure}[p]
  \begin{center}
    \FigureFile(\textwidth,){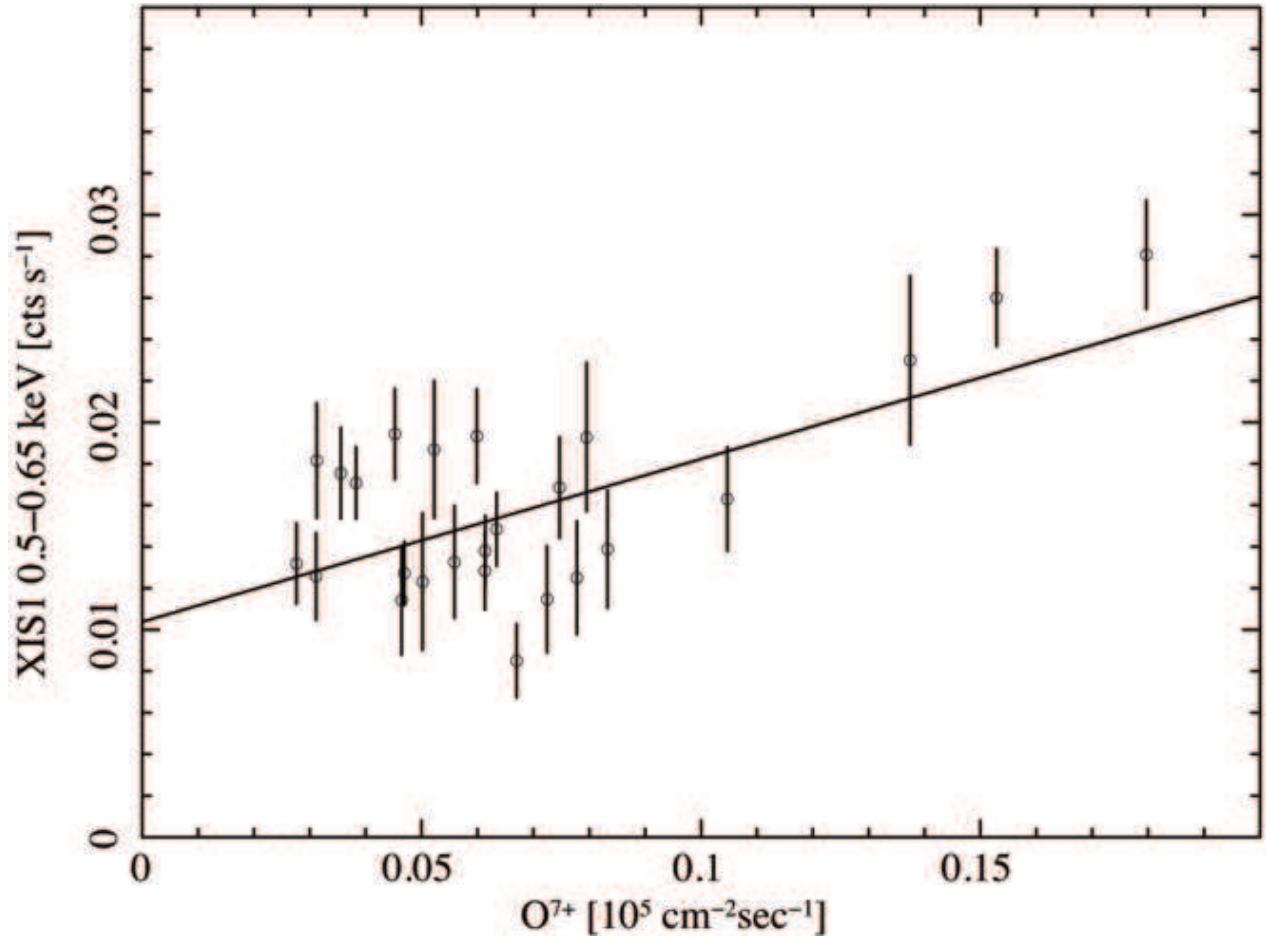}
  \end{center}
  \caption{Correlation between the XIS1 O\emissiontype{VII} line emission 
  and ACE O$^{7+}$ flux curves, considering the 8192 s time 
  delay of the Suzaku data (see text in \S \ref{sec:time}). 
  The vertical error bar is at 1$\sigma$ significance.
  The solid curve is the best-fit linear function.
  }\label{fig:corr}
\end{figure}

In figure \ref{fig:corr}, we plot a relation between the XIS O\emissiontype{VII} line 
rate and the ACE O$^{7+}$ flux considering the time delay of 8192 s. We 
fitted the data with a linear function (solid line).
The best-fit function was expressed as,
\begin{equation}
{\rm {\it C}_{XIS1}~ [cts~ s^{-1}]} = 
{\rm {\it C}_{O^{7+}}~ [10^{5}~ cm^{-2}~ s^{-1}]}\times(7.8\pm1.3)\times10^{-2} 
+ (1.0\pm0.1)\times10^{-2},
\label{eqn:corr}
\end{equation}
where $C_{\rm XIS1}$ and $C_{\rm O^{7+}}$ are the XIS1 0.5--0.65 keV count
rate and the ACE O$^{7+}$ flux, respectively. The error is 1$\sigma$ statistical 
one. The reduced $\chi^2$ was 0.32 for 25 degree of freedom and hence acceptable.
This relation suggests that a linear increase of the XIS1 count rate according to 
the ACE O$^{7+}$ flux, which coincides with the SWCX picture. 
To know the effect of the assumed time delay on the relation, we tested other 
values (0 and 16384 s). We found that, even in these cases, the best-fit linear 
and constant coefficiencies coincide with those assuming 8192 s within 1 $\sigma$ 
fitting errors. 

We noticed that, even if the solar wind O$^{7+}$ flux is zero, there remains a 
certain offset component. This is considered to consist of the instrumental and 
sky background.
To clarify the offset component, we estimated the instrumental background using 
the XIS background data base, which was built from night Earth observations 
\citep{Tawa08}, in the same way as \citet{Ebisawa08}.
The XIS1 0.5--0.65 keV instrumental background was estimated as 
$(3.9\pm0.1)\times10^{-3}$ cts s$^{-1}$ which occupies 40 \% of the 
offset value.
The rest must be a stable background from the sky, $(0.6\pm0.1)\times10^{-2}$ 
cts s$^{-1}$ corresponding to $\sim4$ LU in flux.

The most plausible interpretation for this sky background is a sum of the 
heliospheric SWCX and the soft X-ray background. The former occurs when the
solar wind ions react with a neutral target in the heliosphere (e.g., \cite{Koutroumpa07}) 
but its contribution is unclear, since it is observationally difficult to distinguish 
the heliospheric SWCX from the geocoronal SWCX except by time variations. 
The latter is considered to consist of faint extragalactic sources and 
emission from highly ionized ions in solar neighborhoods and in our Galaxy 
(e.g., \cite{McCammon02, Masui09, Yoshino09}). 
Recently, \citet{Masui09} studied the soft X-ray background from the galactic 
disk ($l=235^\circ,~ b=0^\circ$) with Suzaku. %% From the spectral model, they 
%% concluded that faint dM stars can explain the most of the emission. 
The O\emissiontype{VII} line flux in this line of sight was 2.93$\pm$0.45 LU. 
The line of sight of our Galactic ridge observation ($l=28.463^\circ,~ b=-0.204^\circ$) 
is in the direction of the galactic disk. The slightly different flux may be 
due to possible directional dependence of the soft X-ray background and/or 
seasonal or directional differences in the heliospheric SWCX. 

As an additional background, we also considered the fluorescence 
scattering of solar X-rays by the Earth's atmosphere. The fluorescent 
oxygen line at 0.53 keV is sometimes seen in past Suzaku observations
(e.g., \cite{Miller08}).
The contribution of the fluorescent emission line depends on both 
the amount of the solar X-rays during the observation and the Earth's 
atmosphere density along the line of sight, which is related to the
telescope elevation angle from the Earth (ELV).
To estimate this influence, we filtered the XIS1 data with different 
ELV of $>5^\circ$ (default criteria), $>10^\circ$, $>20^\circ$, 
$>30^\circ$, and $>50^\circ$. 
We then found no significant change in the count rate when the ELV
criteria is different. This suggests a negligible contribution from 
the fluorescence scattering. 
This is consistent with the fact that the solar X-rays were quite 
stable at a low level during the observation ($7\times10^{-7}\sim3\times10^{-8}$ 
W m$^{-2}$ in the GOES12 1.0-8.0 \AA~ data)\footnote{http://www.swpc.noaa.gov/}.
This flux level corresponds to the minimum class solar flare (class A) 
or less. Hence, we concluded that the fluorescent scattering of solar 
X-rays is negligible in our data. 

\section{Discussion}
\label{sec:discuss}

We analyzed the Suzaku XIS1 Galactic ridge observation data and 
discovered a good time correlation between the XIS1 O\emissiontype{VII} count 
rate and the ACE O$^{7+}$ ion flux. From the line center energy 
and the time correlation with the solar wind, we concluded that 
this emission originates from the geocoronal SWCX. 
The average O\emissiontype{VII} line flux was 11 LU, composed of 7 LU from the
geocoronal SWCX and 4 LU from the sky background including the 
soft X-ray background and the heliospheric SWCX. 
Below, we discuss the expected line intensity of the geocoronal 
SWCX.

\begin{figure}[p]
  \begin{center}
    \FigureFile(\textwidth,){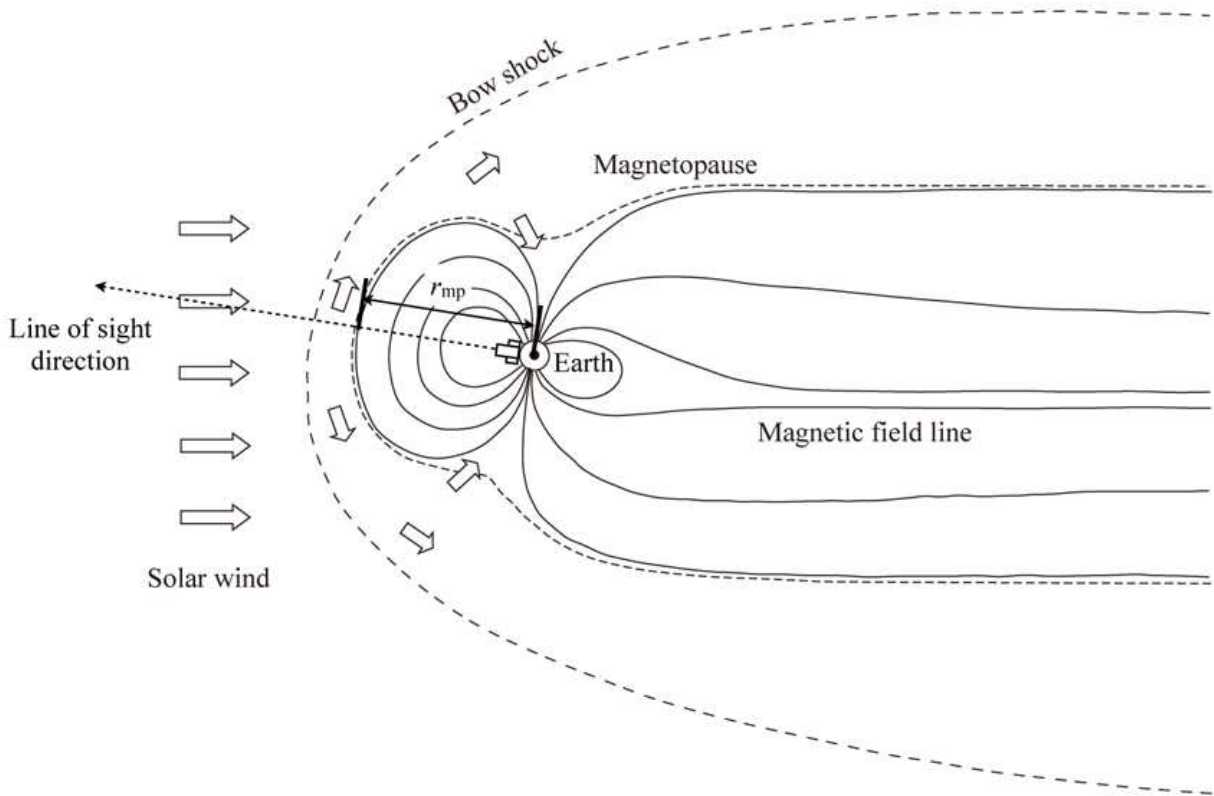}
  \end{center}
  \caption{Schematic view of the magnetosphere and line of 
  sight of the satellite, observed in the GSE x-z plane. 
%  Although the line of sight direction apparently points the 
%  Sun, the average 
  The Sun angle (Sun-Suzaku-object) during the observation 
  was $\sim67^\circ$ and the line of sight has a $+57^\circ$ 
  azimuth angle, perpendicular to this diagram.
  }\label{fig:config}
\end{figure}

Figure \ref{fig:config} shows the schematic view of the Earth's 
magnetosphere and line of sight in our observation. While the 
north ecliptic pole observation with Suzaku was in the direction
of the north pole (Dec$=66^\circ$) \citep{Fujimoto07}, the line
of sight in this observation is almost vertical to it (Dec$=-4^\circ$). 
The line of sight corresponds to the sub-solar side of the 
magnetosheath which is predicted to be a high SWCX flux region
by the theoretical model \citep{Robertson03}.

The line of sight of the Galactic ridge also makes a clear 
difference in the geocentric distance of the point whose 
geomagnetic field becomes open to space for the first time 
along the line of sight ($r_{\rm mp}$, see fig. \ref{fig:config} 
for definition). 
This definition is the same as that in \citet{Fujimoto07}.
In our case, it is equal to the line-of-sight distance 
to the magnetopause boundary which is the location where the 
outward magnetic pressure balances the solar wind pressure. 
Here $r_{\rm mp}$ is an important parameter to estimate the geocoronal
SWCX intensity because most of the solar wind are blocked by the Earth's 
magnetic field and hence interact around $r_{\rm mp}$. Since the Earth's 
exospheric density strongly depends on the distance from the Earth (e.g., 
\cite{Ostgaard03}), $r_{\rm mp}$ should have a large impact on the SWCX intensity.

\citet{Fujimoto07} evaluated $r_{\rm mp}$ using the software
GEOPACK-2005 and T96 magnetic field model \citep{Tsyganenko05}
\footnote{http://modelweb.gsfc.nasa.gov/magnetos/data-based/modeling.html}.
In this model, they took into account solar wind parameters,
that can influence on the structure of the magnetosphere, 
using CDAWeb (Coordinated Data Analysis Web)
\footnote{http://cdaweb.gsfc.nasa.gov/cdaweb/sp phys/}.
We calculated $r_{\rm mp}$ in the same way and found that 
it is almost constant around $\sim$12 $R_{\rm E}$, corresponding
to the distance to the magnetopause. This is natural consequence 
if we consider the line of sight direction (see fig. \ref{fig:config}). 
$r_{\rm mp}$ is larger than that in the north ecliptic pole 
(2$\sim$8 $R_{\rm E}$).
In order to investigate whether the short term variability
seen in the north ecliptic pole was due to the changing $r_{\rm mp}$, 
we checked the short term variability less than several hours
by changing the bin size of the O\emissiontype{VII} light curve from 512 s 
to 8096 s in our data.
Then we found no sign of such a variability, which is 
consistent with the almost constant $r_{\rm mp}$. 

We then proceeded to estimate the expected X-ray line intensity
from an equation below,
\begin{equation}
\displaystyle
f_{\rm OVII} \sim \frac{1}{4\pi}\int^{l_{\rm max}}_{l_{\rm min}} \alpha f_{\rm O7+} n_{\rm H}(l)~ dl~{\rm [photon~ s^{-1}~ cm^{-2}~ str^{-1}]}\\
\label{eqn:swcx}
\end{equation}
where $f_{\rm OVII}$ and $f_{\rm O7+}$ are the O\emissiontype{VII} line intensity and solar wind
O$^{7+}$ flux, $n_{\rm H}$ is the neutral geocoronal hydrogen density, and $l$ 
is the line of sight. $l_{\rm max}$ and $l_{\rm min}$ are the most distant and closest 
distances at which the solar wind ions can interact with the neutral target. 
$\alpha$ contains the atomic cross section and transition probability information.

To calculate $f_{\rm OVII}$, we made simple assumptions as below. We used 
$\alpha$ of 6$\times10^{-15}$ cm$^{-2}$ from \citet{Wegmann98}, assuming that 
all transitions are equally probable. The Earth's exospheric neutral hydrogen 
density model by \citet{Ostgaard03} was used for $n_{\rm H}(l)$. 
Because most of the solar ions will be trapped near the Earth's magtopause
in the direction of the Galactic ridge, we simply assumed $l_{\rm min}=r_{\rm mp}$.
$l_{\rm max}$ of 20 $R_{\rm E}$ was assumed as a rough estimate, because
$n_{\rm H}(l)$ at $>$12 $R_{\rm E}$ is not well known due to the very low 
density ($<20$ cm$^{-3}$, see fig. 10 in \cite{Ostgaard03}). 
The average ACE O$^{7+}$ ion flux of $\sim1\times10^{-5}$ cm$^{-2}$ s$^{-1}$ 
was used as $f_{\rm O7+}$. 
Then, the O\emissiontype{VII} line flux was estimated as 0.2 LU, 
which was 35 times smaller than the observation. Even if we integrate the
equation up to 100 $R_{\rm E}$, it was 0.3 LU, still 20 times smaller.

In the above calculation, we did not take into account an effect of the bow 
shock across which can increase the solar wind flux.
Then, we tried to take into account this effect based on a standard
theoretical model given by \citet{Spreiter66}. According to their calculation, 
the solar wind mass flux increases 2$\sim$3 times across 
the bow shock in the sub-solar side. Although we take into account this effect,
there remains a factor of $\sim$10 discrepancy between the observed and 
expected O\emissiontype{VII} line flux.
This uncertainty can be due to an uncertainty of the neutral hydrogen model 
at $>$10 $R_{\rm E}$ and/or the solar wind distribution.

The time delay between the X-ray light curve and the solar wind ion flux
can provide us with information on the propagation of the solar wind. 
\citet{Carter10} checked the ACE and WIND proton curves, and estimated
the orientation of the solar wind wavefront assuming a planar wavefront. 
In our case, a part of the ACE proton flux during the time variation is 
unfortunately unavailable. On the other hand, as we described in \S 
\ref{sec:time}, the time delay between the Suzaku light curve and the ACE 
ion flux of 0$\sim$16384 sec is consistent with the calculation assuming 
the planar wavefront and constant solar wind velocity. This supports that 
enhancement of the geocoronal SWCX diffuse X-ray background can be predicted 
and excluded if there are simultaneous solar wind proton and ion observations.

Finally, we point out that the geocoronal SWCX in combination with the 
solar wind data can be a useful tool to evaluate the Earth's exospheric 
density model and the solar wind distribution. 
Our observation had a line of sight which intersects with the sub-solar 
side of the magnetosheath, which is predicted to have a high SWCX flux. 
Archival Suzaku data analysis will enable us to examine the directional 
dependence even more. Our observation demonstrated that the time delay 
and intensity relationship between the X-ray O\emissiontype{VII} line and solar wind 
O$^{7+}$ flux must be a key to investigate the line of sight dependence 
of the geocoronal SWCX.
Future high-energy resolution observations with an X-ray 
microcalorimeter onboard Astro-H will make it possible to monitor
the conditions of the Earth's exosphere and solar wind more precisely. 

The authors acknowledge the Suzaku XIS instrument and operation team, 
and the ACE SWEPAM/SWICS instrument team and the ACE Science Center.

\clearpage

%\begin{figure}[p]
%  \begin{center}
%    \FigureFile(0.7\textwidth,){fig1av1p.eps}
%  \end{center}
%  \caption{
%  }\label{fig:test}
%\end{figure}

\end{document}